\documentclass[12pt,a4paper]{article}
\usepackage{amssymb}
\usepackage{epsf}
\let\tsection\section
\renewcommand{\section}{\setcounter{equation}{0}\tsection}

\def\bbr{\mathbb R}
\def\bbz{\mathbb Z}
\def\tr{\mathop{\rm Tr}}
\def\Prob{\mathop{\rm Prob}\nolimits}
\def\ts{\textstyle}
\def\Pt{\tilde{\cal P}}
\def\M{{\cal M}}\def\F{{\cal F}}

\def\tauul{\underline\tau}
\def\iul{\underline i}\def\xul{\underline x}
\def\piul{\underline\pi}
\def\nul{\underline n}
\def\ts{\textstyle}
\def\P{{\cal P}}
\def\mubar{\bar\mu}\def\nubar{\bar\nu}

\begin{document}

\begin{center} 
{\bf \Large Entropy of Open Lattice Systems}
\vskip20pt

B. Derrida\footnote{Laboratoire de Physique Statistique,
Ecole Normale Sup\'erieure, 24 rue Lhomond, 75005 Paris, France; 
email derrida@lps.ens.fr.},
J. L. Lebowitz\footnote{Department of Mathematics,
Rutgers University, New Brunswick, NJ 08903; email lebowitz@math.rutgers.edu,
speer@math.rutgers.edu.}${}^{\rm ,}$\footnote{Also Department of Physics, 
Rutgers.},  and E. R. Speer${}^{\scriptstyle 2}$

\end{center}

\vskip20pt 
\noindent 
{\bf Abstract:} We investigate the behavior of the Gibbs-Shannon entropy of
the stationary nonequilibrium measure describing a one-dimensional lattice
gas, of $L$ sites, with symmetric exclusion dynamics and in contact with
particle reservoirs at different densities.  In the hydrodynamic
scaling limit, $L\to\infty$, the leading order ($O(L)$) behavior of this
entropy has been shown by Bahadoran to be that of a product measure
corresponding to strict local equilibrium; we compute the first correction,
which is $O(1)$.  The computation uses a formal expansion of the entropy in
terms of truncated correlation functions; for this system the $k^{\rm th}$
such correlation is shown to be $O(L^{-k+1})$.  This entropy correction
depends only on the scaled truncated pair correlation, which describes the
covariance of the density field.  It coincides, in the large $L$ limit, with
the corresponding correction obtained from a Gaussian measure with the same
covariance.

\section{Introduction}

The properties of nonequilibrium stationary states (NESS) of open systems,
i.e., ones in contact with infinite reservoirs at different chemical
potentials and/or temperatures, is a subject of great interest
\cite{KMP}--\cite{EY}. The simplest nontrivial example of such a system
is the one-dimensional simple symmetric exclusion processes (SSEP) on the
finite lattice $\Lambda_L=\{1,2,\ldots,L\}$, with particle reservoirs
coupled to sites $1$ and $L$; we take these reservoirs to have densities
$\rho_a$ and $\rho_b$, respectively, with $\rho_a>\rho_b$.  The $2^L$
possible configurations of the system are described by the $L$-tuple
$\tauul_L=(\tau_1,...,\tau_L)$, with $\tau_i=1$ if site $i$ is occupied and
$\tau_i=0$ if the site is empty.  The stationary measure
$\bar \mu_L(\tauul_L)$ of the system is explicitly known in terms of
products of noncommuting matrices \cite{DEHP,LDF}.  Using this
representation it is possible to obtain considerable information about the
truncated correlation functions at $k$ distinct sites,
$\langle\tau_{i_1}\cdots\tau_{i_k}\rangle^T_{\bar\mu_L}$.  In particular
these are $O(L^{-k+1})$, in the sense that
 \begin{equation}
\lim_{L\to\infty}L^{k-1}\langle\tau_{\lfloor x_1L\rfloor}\cdots
  \tau_{\lfloor x_kL\rfloor}\rangle^T_{\mubar_L}
    =F_k(x_1,\ldots,x_k),
  \label{Fk}
 \end{equation}
 for certain continuous functions $F_1(x_1)$, $F_2(x_1,x_2)$, etc; here
$\lfloor \xi\rfloor$ is the greatest integer not exceeding $\xi$.  These
correlations are thus very long range and contribute, despite their
vanishing pointwise as $L\to\infty$, to the fluctuations (and larger
deviations) about the typical density profile
$\bar\rho(x)=F_1(x)=\lim_{L\to\infty}\langle\tau_{\lfloor
xL\rfloor}\rangle$ in the hydrodynamic scaling limit, $L\to\infty$,
$i/L\to x\in[0,1]$, where $i=1,\ldots,L$ labels the lattice sites
\cite{LDF}.

In this hydrodynamic limit the typical density profile $\bar \rho(x)$ is
the stationary solution of the macroscopic hydrodynamic equation with
boundary conditions $\bar\rho(0)=\rho_a$, $\bar\rho(1)=\rho_b$.
For the SSEP this is the simple diffusion equation and so 
 \begin{equation}
\bar\rho(x)=\rho_a(1-x)+\rho_bx.  \label{rhobar}
 \end{equation}
 The fluctuations about the typical
profile $\bar \rho(x)$ are given by a Gaussian field whose covariance is
determined by the truncated pair correlation function \cite{S2,LDF}.  The
result agrees with that obtained from fluctuating hydrodynamics \cite{S}.

In this paper we study the relation between the functions $F_k$ and
the $L\to\infty$ limit of the Gibbs-Shannon entropy of the stationary
measure $\bar\mu_L$ , defined for any measure $\mu_L$ by 
 \begin{equation}
 S(\mu_L)=-\sum_{\tauul_L}\mu_L(\tauul_L)\log\mu_L(\tauul_L).\label{entropy}
 \end{equation}
 In this limit $S(\bar\mu_L)$ is $O(L)$ and 
only $F_1$ is relevant to leading order (this is a result of
Bahadoran \cite{Bahadoran}) and our goal here is to show that only $F_1$
and $F_2$ are relevant for the first correction, which is $O(1)$.  

Our motivation for studying $S(\bar\mu_L)$ is its potential connection with
deviations from the typical profile $\bar \rho(x)$ \cite{Var}.  The
expectation of such a connection comes from our experience with equilibrium
systems, for which the probability of such deviations is determined by the
induced change in the entropy.

In fact, the open SSEP with NESS measure $\mubar_L$ is, in the hydrodynamic
scaling limit, very closely related to such a (local) equilibrium system.
To make this more precise, let us define the product measure with expected
density $n_i$ at site $i$ by
 \begin{equation}
 \nu^{(\nul)}_L({\tauul_L}) = \prod_{i=1}^L \,[n_i \tau_i + (1 -
  n_i)(1 - \tau_i)], \label{22}
 \end{equation}
 and for any macroscopic density profile $\rho(x)$ write (with some abuse
of notation) $\nu_L^{(\rho)}=\nu_L^{(\nul)}$ with $n_i=\rho(i/(L+1))$ (the
specific definition of $n_i$ arises from the convention that the system has
total length $L+1$, with the boundary
reservoirs located on sites $i=0$ and $i=L+1$).
 Then the restriction of the NESS measure $\bar\mu_L$ to the
variables $\tau_i$ for $i$ lying in an interval
$\bigl[\lfloor xL\rfloor,\lfloor xL\rfloor+m]$, where $m$ is independent of
$L$, is indistinguishable, for $L\to\infty$, from the corresponding
restriction of the local equilibrium product measure $\nu_L^{(\bar\rho)}$
(see (\ref{rhobar})).

For this  product measure $\nu_L^{(\bar\rho)}$ 
the probability of observing a
density profile $\rho(x)$ is given, for large $L$, by
 \begin{equation}
\Prob(\{\rho(x)\}|\nu_L^{(\bar\rho)}) 
   \sim e^{-L\F_{\rm eq}(\{\rho(x)\})},\label{LDF}
 \end{equation}
 where the free energy or large deviation functional (LDF) $\F_{\rm eq}$
can be written as
 \begin{eqnarray}
   -{\cal F}_{\rm eq}(\{\rho(x)\}) 
  &=& \int_0^1\bigl\{[s(\rho(x))+\lambda(\bar\rho(x))\rho(x)]\nonumber\\
     &&\hskip40pt 
   -[s(\bar\rho(x))+\lambda(\bar\rho(x))\bar\rho(x)]\bigr\}\,dx.
    \label{eqentt}
 \end{eqnarray}
 Here $s(r)=-(r\log r+(1-r)\log(1-r))$ is the entropy per unit length (or
site) of the product measure with constant density $r$, and
 \begin{equation}
\lambda(r)=-{\partial s\over\partial r}(r) =\log\left(r\over1-r\right)
\label{lambda}
 \end{equation}
 is the chemical potential which yields this density.  The connection
 between the LDF $\F_{\rm eq}$ and entropy given in (\ref{eqentt}) extends
 to more general (non product) local equilibrium measures \cite{Olla,Ellis}.

Given this connection between entropy and large deviations in equilibrium
systems, it is natural to ask whether there exists a similar relation
between $S(\bar \mu_L)$ and the large deviation functional in the NESS of
the SSEP, for which $\Prob(\{\rho(x)\}|\bar\mu_L)$ is qualitatively
different from (\ref{LDF}) \cite{DLS0,BDGJL,BDGJL2,LDF}.  Using results of
Kosygina \cite{Kosygina}, Bahadoran showed, for a large class of systems
including the open SSEP, that
 \begin{equation}
\lim_{L \to \infty} \frac{1}{L} S(\bar \mu_L) = \lim_{L \to \infty}
\frac{1}{L} S(\nu_L^{(\bar \rho)}).  \label{Baha}
 \end{equation}
 In other words, the Gibbs-Shannon entropy is, to leading order, exactly
the same as that of the product measure with density $\bar\rho(x)$, i.e.,
$\nu_L^{(\bar \rho)}$.  It thus does not reflect at all the very
different nature of the large deviation functional for the NESS in
comparison with that of equilibrium systems.

Information about the  
probabilities of untypical configurations in the NESS
of the SSEP is encoded in the truncated correlation functions, or
equivalently in the $F_k$'s of (\ref{Fk}). These also contribute to the
entropy $S(\bar\mu_L)$ beyond the leading order.  This is what we
investigate in the present note.  We find that the difference
 \begin{equation}
   R_L\equiv S(\bar \mu_L) - S(\nu_L^{(\bar \rho)}) \label{rldef}
 \end{equation}
  approaches a constant
value $R$ as $L \to \infty$; Bahadoran's theorem only says that it grows
slower than $L$.  Furthermore $R$ depends only on the pair correlation
function $F_2$, indicating that only configurations which contribute to the
Gaussian fluctuations about $\bar \rho(x)$ contribute to the entropy at
this order.  This permits us to obtain an explicit expression for $R$ as
the $L\to\infty$ limit of $\hat R_L$, the corresponding difference in
entropies for a Gaussian measure on variables $\xi_i \in {\mathbb R}$,
$i=1,...,L$, having the same covariance matrix as $\bar \mu_L$.  We present
analytic arguments in favor of this expression, and also check it
numerically via exact computations on systems of different sizes.  It
appears in fact that our results extend to more general systems having long
range truncated correlation functions of the form (\ref{Fk}), as we discuss
in section~\ref{others}.

The outline of the rest of this paper is as follows.  In
section~\ref{models} we describe the SSEP and the corresponding Gaussian
model, and in section~\ref{others} the possible extension to other models,
in particular, the weakly asymmetric exclusion process (WASEP).  In
section~\ref{numbers} we report on numerical computations of the entropy
difference $R_L-\hat R_L$ for different system sizes and densities, in both
the SSEP and the WASEP.  In section~\ref{gaussian} we compute rigorously
the $L\to\infty$ limit of $\hat R_L$ for the Gaussian model.  In
section~\ref{probs} we establish a relation between an arbitrary measure
$\mu(\tauul_\Lambda)$, where $\tauul_\Lambda=(\tau_i)_{i\in\Lambda}$ with
$\Lambda$ any finite set of points, and the truncated correlations
$\langle\prod_{i\in\Lambda'}\tau_i\rangle^T_{\mu_\Lambda}$ for
$\Lambda'\subset\Lambda$; we develop from this an expression for the
entropy in terms of the truncated correlations.  We use this in
section~\ref{particles} to argue that the difference $\lim_{L\to\infty}R_L$
exists and has value $R=\lim_{L\to\infty}\hat R_L$.

\section{The models and the results\label{models}} 

We begin with a full description of the SSEP.  In this model each particle
independently attempts to jump to its right neighboring site, and to its
left neighboring site, in each case at rate $1$ (so that there is no
preferred direction).  It succeeds if the target site is empty; otherwise
nothing happens.  A particle is added to site $1$, when the site is empty,
at rate $\alpha$, and removed, when the site is occupied, at rate $\gamma$;
similarly particles are added to site $L$ at rate $\delta$ and removed at
rate $\beta$.  This corresponds \cite{LDF} to the system being in contact
with infinite left and right reservoirs having respective densities
 \begin{equation}
 \rho_a={\alpha\over\gamma+\alpha}, \qquad \qquad
  \rho_b={\delta\over\beta+\delta}\;. \label{rho12}
 \end{equation}
 We also introduce the parameters
 \begin{equation}
   a={1\over\gamma+\alpha}, \qquad\qquad b={1\over\beta+\delta}.\label{abdef}
 \end{equation}

We give in Appendix A a proof of the scaling form (\ref{Fk}) for this
model. The first three truncated correlations are, for $i<j<l$, 
 \begin{eqnarray}
  \langle\tau_i\rangle_{\mubar_L}
  &=&{\rho_a(L+b-i)+\rho_b(i+a-1)\over L+a+b-1},\label{tau1}\\
  \langle\tau_i\tau_j\rangle^T_{\mubar_L}
   &=&-{(\rho_a-\rho_b)^2(i+a-1)(L+b-j)\over(L+a+b-1)^2(L+a+b-2)},
    \label{tau2}\\
  \langle\tau_i\tau_j\tau_l\rangle^T_{\mubar_L}
 &=&-2{(\rho_a-\rho_b)^3(i+a-1)(L+1+b-a-2j)(L+b-l)\over
     (L+a+b-3)(L+a+b-2)(L+a+b-1)^3}.\ \ \ \ \ \label{tau3}
\end{eqnarray}
 Thus (\ref{Fk}) holds for $k=1,2,3$, where for $x<y<z$, 
 \begin{eqnarray}
 F_1(x)&=&\rho_a(1-x)+\rho_bx,\label{F1}\\
 F_2(x,y)&=&-(\rho_a-\rho_b)^2x(1-y),\label{F2}\\
 F_3(x,y,z)&=&-2(\rho_a-\rho_b)^3x(1-2y)(1-z).\label{F3}
 \end{eqnarray}

We remark that if $a=b=1$ then
$\langle\tau_i\rangle_{\mubar_L}=\bar\rho(i/(L+1))$ and the entropy
difference $R_L$ of (\ref{rldef}) must be negative, since
$\nu_L^{(\bar\rho)}$ maximizes the entropy $S(\mu)$ among all measures
$\mu$ satisfying $\langle\tau_i\rangle_\mu=\bar\rho(i/(L+1))$.  Because our
expression (\ref{FD}) for $\lim_{L\to\infty}R_L$ is independent of $a$ and
$b$, this limit must be negative or zero.

  In the remainder of the paper we argue that, in the SSEP, the
next order correction to the result of Bahadoran will be equal to the
correction for a Gaussian system with the same covariance.  Specifically,
let $\hat\nu_L$ and $\hat\mu_L$ be Gaussian measures on $L$ variables with
mean zero and respective covariance matrices $J_L$ and $K_L$ given by
 \begin{eqnarray}
   (J_L)_{ii} 
    =\langle \tau_i\rangle_{\mubar_L}(1-\langle \tau_i\rangle_{\mubar_L}),
   &\qquad& (J_L)_{ij}=0,\quad i\ne j;\\
   (K_L)_{ii} =\langle \tau_i\rangle_{\mubar_L}(1-\langle \tau_i\rangle_{\mubar_L}),
   &\qquad& (K_L)_{ij}=\langle \tau_i\tau_j\rangle^T_{\mubar_L},\quad i\ne j.
 \end{eqnarray}
  We note from (\ref{tau1}) that $\langle \tau_i\rangle_{\mubar_L}$ and
$1-\langle \tau_i\rangle_{\mubar_L}$ do not vanish.  The entropies of these
Gaussian measures are given by
 \begin{eqnarray}
   S(\hat\nu_L) &=& {L\over 2}(1+\log 2\pi)+{1\over 2}\log \det J_L,\\
   S(\hat\mu_L) &=& {L\over 2}(1+\log 2\pi)+{1\over 2}\log \det K_L,
 \end{eqnarray}
 so that 
 \begin{equation}
   \hat R_L\equiv S(\hat\mu_L)- S(\hat\nu_L) 
    = {1\over 2}\log {\det K_L \over \det J_L}.
  \label{Sg}
 \end{equation}
 The $L\to\infty$ limit of (\ref{Sg}),
 \begin{equation}
 R  =   \lim_{L\to\infty} \hat R_L
  = \lim_{L\to\infty}{1\over 2}\log {\det K_L \over \det J_L},
   \label{limits}
 \end{equation}
  exists; see section~\ref{gaussian}.  Further, we claim that this limit
gives also the lowest order correction to the result (\ref{Baha}) of
Bahadoran (see (\ref{rldef})) :
 \begin{equation}
   \lim_{L\to\infty} R_L 
    =\lim_{L\to\infty}[S(\bar \mu_L) - S(\nu_L^{(\bar \rho)})]
    = R.\label{limits2}
 \end{equation}

\subsection{Other models\label{others}}

It is natural to ask to what extent these results hold for other lattice
gas models.  As will become clear in section~\ref{particles}, the key
element in our analysis for the SSEP is the scaling behavior (\ref{Fk}) of
the truncated correlation functions.  (We also use some technical facts
about the way the limit in (\ref{Fk}) is achieved and the size of the
limiting functions $F_k$.)  We expect (\ref{limits2}) to hold for models
having this same scaling behavior and satisfying an additional condition
discussed in Remark 4.1.

Unfortunately, less is known about the correlation functions for other
lattice gas models than for the SSEP; in particular, we know of no other
model of an open NESS for which (\ref{Fk}) has been established with
nonzero $F_k$, $k\ge2$ (for zero range processes the NESS is a product
measure, i.e., $F_k=0$ for $k\ge2$).  We expect, however, that this scaling
will hold in the weakly asymmetric simple exclusion process (WASEP); see
\cite{WASEP}, where expressions corresponding to (\ref{F1}) and (\ref{F2})
are given for this model.  In the WASEP the boundary dynamics are those of
the SSEP, but the bulk dynamics are modified so that a particle attempts to
hop to its right at rate 1 and to its left at rate $\exp(-\lambda/L)$;
$\lambda$ is a parameter which interpolates between the symmetric process
($\lambda=0$) and the totally asymmetric process ($\lambda=\pm \infty$).
The typical profile $\bar\rho(x)$ is the solution of the viscous Burgers
equation.  We include numerical results for the WASEP in the next section.

The truncated correlation functions are also expected to satisfy (\ref{Fk})
in the KMP model \cite{KMP}.  This is an open system in which the
variable $\xi_i\in\bbr_+$ at site $i$, $i=1,\ldots,L$, represents an energy
at that site, and $\rho_a$ and $\rho_b$ are replaced by temperatures $T_a$
and $T_b$.  For this system \cite{KMP},
 \begin{equation}
  F_1^{\rm KMP}(x)
  =\lim_{L\to\infty} \langle\xi_{\lfloor xL\rfloor}\rangle
   =T_a(1-x)+T_bx,
 \end{equation}
  while for $x<y$ \cite{BGL},
 \begin{equation}
  F_2^{\rm KMP}(x,y)=\lim_{L\to\infty} 
   L\langle\xi_{\lfloor xL\rfloor}\xi_{\lfloor yL\rfloor)}\rangle
   = (T_a-T_b)^2x(1-y).
 \end{equation}

One can also imagine that for more general diffusive systems, such as those
described by the macroscopic fluctuation theory \cite{BDGJL,BDGJL2}, the
long range part of the truncated correlation functions scales as in (1.1)

\section{Numerical results\label{numbers}}

We have investigated (\ref{limits2}) numerically for the SSEP, at several
different values of the boundary densities $\rho_a,\rho_b$, and for the
WASEP, at $\rho_a=1$, $\rho_b=0$, for several different values of
$\lambda$.  For all computations we have taken $a=b=1$ (see
(\ref{abdef})).  We were able to consider systems up to size $L = 25$.  In
each case we computed the measure $\mubar_L$ explicitly and from this
$S(\mubar_L)$, $S(\nu_L^{(\rho)})$, $S(\hat\mu_L)$, and $S(\hat\nu_L)$, and thus,
from (\ref{rldef}) and (\ref{Sg}), $R_L$ and $\hat R_L$.  For the SSEP we
could also compute the limiting value $R$, defined in (\ref{limits}), to a
high degree of accuracy, using (\ref{tau2}).

Figures 1 and 2 present our results for the SSEP. In order to show
results for several parameter values on the same figure, we plot the
normalized difference
 \begin{equation}
{R_L-\hat R_L\over R}
 \end{equation}
 as a function of $1/L$.  The table within each figure gives the values of
$\rho_a$ and $\rho_b$ for each curve, as well as the corresponding value
of $R$.  Confirmation of (\ref{limits2}) corresponds in each case to
$\lim_{1/L\to0}(R_L-\hat R_L)/R=0$.  This certainly appears to hold, but the
maximum system size we have been able to achieve is perhaps too small for
the evidence to be completely convincing.

\begin{figure} [!t]
\centerline{\epsffile{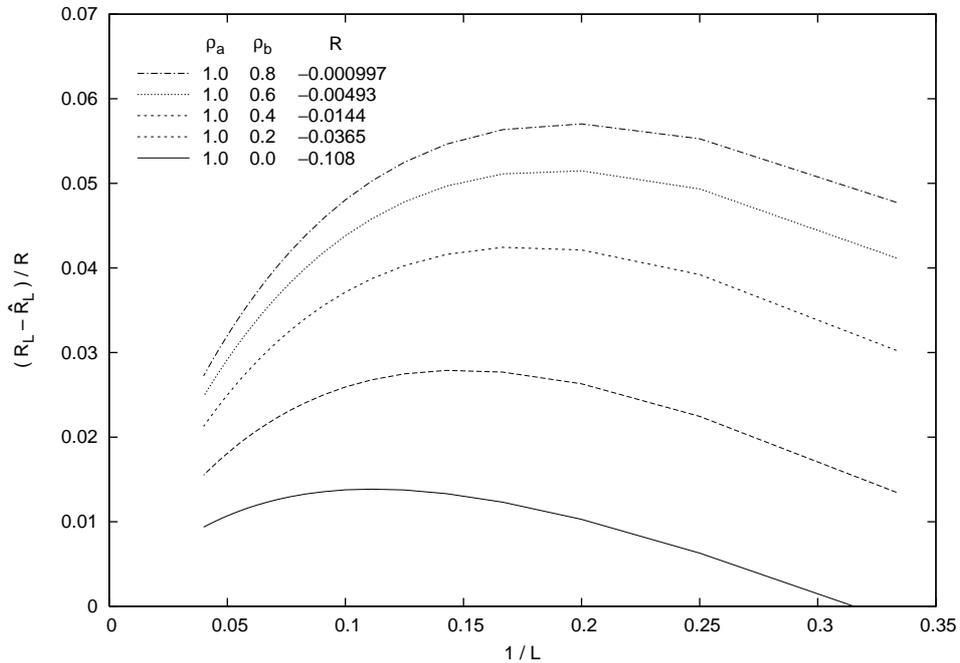}}
 
\caption{Differences of corrections to entropies in the SSEP and Gaussian
models for $\rho_a=1$ and several choices of $\rho_b$.  The data are
consistent with the vanishing of $R_L-\hat R_L$ as $L\to\infty$.}

\end{figure}
  
\begin{figure} [!t]
\centerline{\epsffile{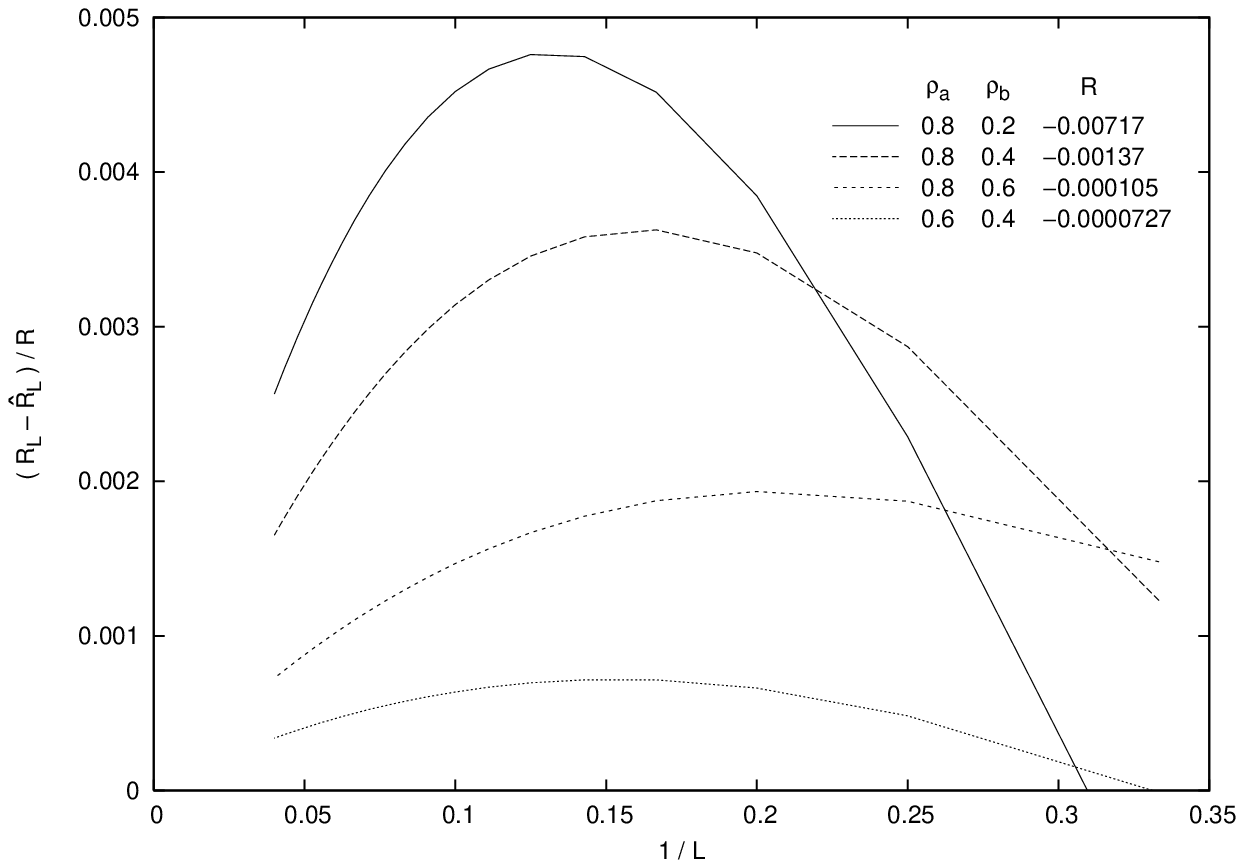}}
 \caption{Same as Figure 1 for other choices of $\rho_a$ and $\rho_b$.}
\end{figure}
  
Figure 3 gives similar plots for the WASEP, at different values of
$\lambda$, with $\rho_a=1$, $\rho_b=0$.  Here we have no closed form for
the two-point correlation function, so that an accurate computation of $R$
is more difficult than for the SSEP; we therefore plot the unnormalized
difference $R_L-\hat R_L$ against $1/L$.  The behavior for small $L$ is
quite irregular, particularly for negative values of $\lambda$, 
but the large-$L$ behavior again provides some
confirmation that $\lim_{1/L\to0}(R_L-\hat R_L)=0$, i.e., that
(\ref{limits2}) holds for this model.

\begin{figure} [!tb]
\centerline{\epsffile{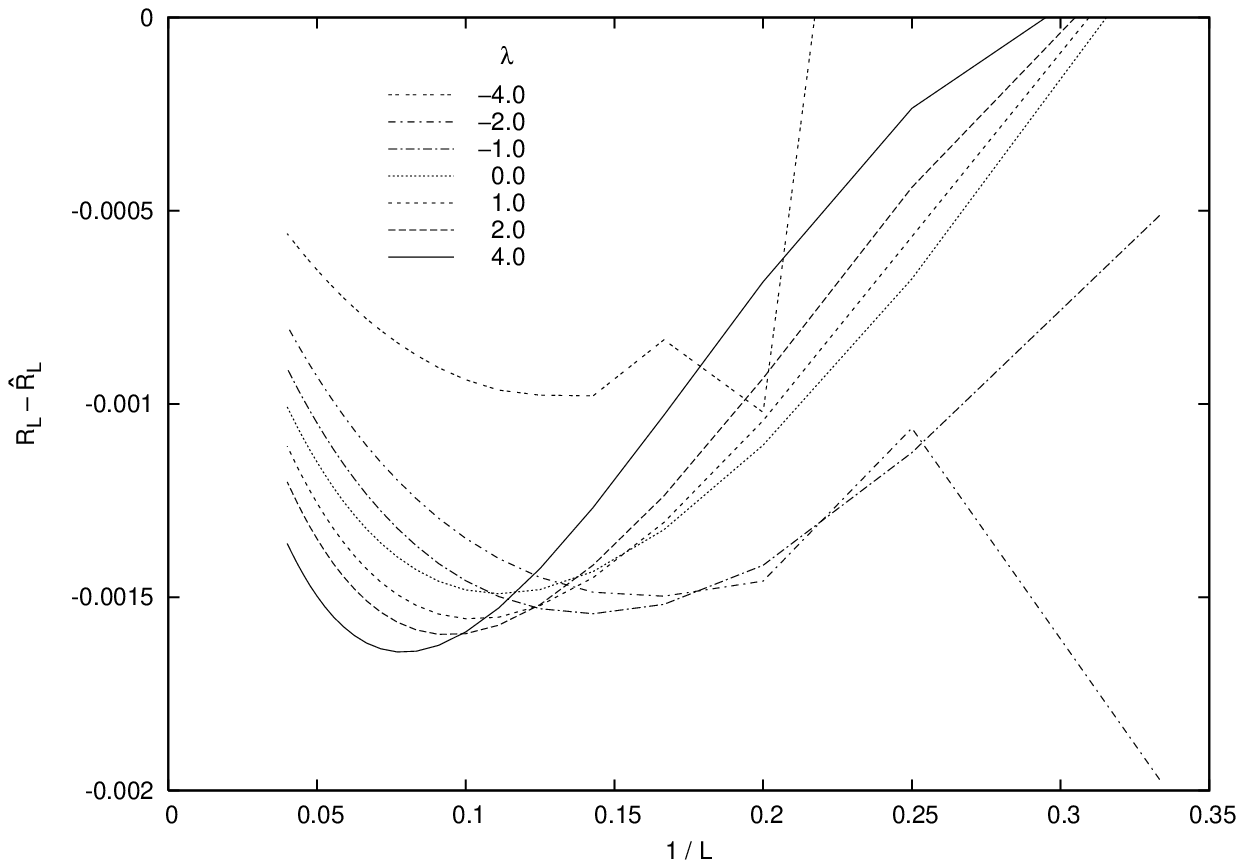}}
 \caption{Same as Figures 1 and 2 for the WASEP.}
\end{figure}

\section{The Gaussian limit\label{gaussian}} 

In this section we evaluate the limit $R$ of (\ref{limits}).  Let us write
 \begin{equation}
{\det K_L \over \det J_L}
  =\det \left[J_L^{-1/2}\, K_L \,J_L^{-1/2}\right]
  = \det(I+U_L), \label{detU}
 \end{equation}
 where
 \begin{equation} 
  U_L=J_L^{-1/2}[K_L-J_L]J_L^{-1/2},
    \end{equation}
 so that $(U_L)_{ii}=0$, $i=1,\ldots,L$, and  
 \begin{equation}
 (U_L)_{ij}={\langle \tau_i\tau_j\rangle^T \over
  \sqrt{\langle \tau_i\rangle(1-\langle \tau_i\rangle)
   \langle \tau_j\rangle(1-\langle \tau_j\rangle)}},
   \qquad 1\le i\ne j\le L.\label{Uoffdiag}
 \end{equation}
 In order to pass to a continuum limit it is convenient to relate
$U$ to the integral operator $H_L$ on $L^2([0,1])$ with  kernel
 \begin{equation}
h_L(x,y) = L(U_L)_{ij},\quad
   \hbox{for}\ {i-1\over L}<x\le{i\over L},\ {j-1\over L}<y\le{j\over L},\ 
 \label{HLdef}
 \end{equation}
 that is, $(H_L\phi)(x)=\int_0^1h_L(x,y)\phi(y)\,dy$ for $\phi\in L^2([0,1])$.
Since $H_L$ has  rank (at most) $L$, all but $L$  of the eigenvalues of
$I+H_L$ are equal to 1, so that the determinant $\det(I+H_L)$ is certainly
well defined.  Then
 \begin{equation}
  \det(I+U_L)=\det(I+H_L), \label{deteq}
 \end{equation}
 since if we define
 \begin{equation}
\psi_{L,i}(x)=\cases{\sqrt L,& if $(i-1)/L<x\le i/L$,\cr0,&otherwise,\cr}
 \end{equation}
 then the $\psi_{L,i}$ for $i=1,\ldots,L$ form an orthonormal set in
$L^2([0,1])$ which spans the range of $H_L$ and satisfies
$H_L\psi_{L,i}=\sum_j(U_L)_{ji}\psi_{L,j}$.

  Now (\ref{Fk}) implies that for $x\ne y$,
$\lim_{L\to\infty}h_L(x,y)=h(x,y)$, where
 \begin{equation}
  h(x,y)={F_2(x,y)\over \sqrt{F_1(x)(1-F_1(x))F_1(y)(1-F_1(y))}}.\label{Fred}
 \end{equation}
  Let $H$ be the integral operator on $L^2([0,1])$ with kernel $h$.  It can
be shown that $H$ is of trace class (we define this
precisely below), so that $\det(I+H)$ is well defined \cite{Simon}.
However, it is not true that $\lim_{L\to\infty}\det(I+H_L)=\det(I+H)$;
essentially, this is because the diagonal elements of $H_L$ are zero rather
than being given by the obvious extension of (\ref{Uoffdiag}), and as a
consequence $H_L$ does not converge to $H$ in trace norm.  To evaluate the
limit correctly it is helpful to introduce the {\sl regularized
determinant} \cite{Simon}; one needs then only convergence of $H_L$ to $H$
in a weaker sense.

 We now discuss the general theory of the 
 regularized determinant in the (relatively simple)
context in which we will use it. Let $A$ be a compact integral operator on
$L^2([0,1])$ with symmetric kernel $a(x,y)=a(y,x)$, so that
$(A\phi)(x)=\int_0^1a(x,y)\phi(y)\,dy$; $A$ is self-adjoint and hence
diagonalizable: $A\phi_n=\lambda_n\phi_n$ for some orthonormal basis
$\phi_n$. $A$ is of {\sl trace class} if
$\|A\|_1\equiv\sum_n|\lambda_n|<\infty$ and of {\sl Hilbert-Schmidt class}
if $\|A\|_2\equiv\sum_n|\lambda_n|^2<\infty$; we also have
 \begin{equation}
\|A\|_2\equiv\int_0^1\int_0^1|a(x,y)|^2\,dx\,dy.
 \end{equation}
 If $A$ is of trace class then both the trace $\tr A=\sum_n\lambda_n$ and
the Fredholm determinant $\det(I+A)=\prod(1+\lambda_n)$ are well defined
and satisfy $\log\det(I+A)=\tr\log(I+A)$.  If $A$ is of
Hilbert-Schmidt class then $\det(I+A)$ may not be defined but
$\tilde A=e^{-A}(I+A)-I$ is of trace class and the regularized determinant
$\det_2(I+A)$ is defined by
 \begin{equation}
\det\nolimits_2(I+A)=\det(I+\tilde A).\label{det2}
 \end{equation}
 We note several properties of $\det_2$ which we will need below: (i)~if
$A$ is of trace class then
 \begin{equation}
\det\nolimits_2(I+A)=\det(I+A)e^{-\tr A}; \label{detdet2}
 \end{equation}
  (ii)~if $A$ is Hilbert-Schmidt and $k\ge2$ then $A^k$ is of trace class,
with
 \begin{equation}
  \tr  A^k = \int_0^1dx_1\cdots\int_0^1 dx_k\;
         h(x_1,x_2)\cdots h(x_{k-1},x_k)h(x_k,x_1),\label{trint}
 \end{equation}
 and $\tr A^k$ is continuous in the Hilbert-Schmidt norm, i.e.,
 \begin{equation}
\lim_{n\to\infty}\tr A_n^k  =\tr A^k
   \quad\hbox{if}\quad   \lim_{n\to\infty}\|A_n-A\|_2=0;\label{conttr}
\end{equation}
  (iii)~$\det_2(I+A)$ is continuous in the Hilbert-Schmidt norm, i.e.,
 \begin{equation}
\lim_{n\to\infty}\det\nolimits_2(I+A_n)=\det\nolimits_2(I+A) 
   \quad\hbox{if}\quad   \lim_{n\to\infty}\|A_n-A\|_2=0;\label{contdet}
\end{equation}
  and (iv)~if the operator
norm of $A$ satisfies $\|A\|<1$ then
 \begin{equation}
\log\det\nolimits_2(I+A)=\sum_{n=2}^\infty{(-1)^{n+1}\over n}\tr A^n.
 \label{P3}
 \end{equation}

  In order to apply these ideas we note that it follows from (\ref{tau1}),
(\ref{tau2}), (\ref{F1}), and (\ref{F2}) that the kernel $h(x,y)$ is square
integrable and that
 \begin{equation}
\lim_{L\to\infty}\|H_L-H\|_2
  =\lim_{L\to\infty}\int_0^1\int_0^1|h_L(x,y)-h(x,y)|^2\,dx\,dy =0.
 \label{Hcont}
 \end{equation}
 Now from (\ref{detdet2}) and the fact that $\tr H_L=0$ it follows that
$\det(I+H_L)=\det_2(I+H_L)$, and hence from (\ref{detU}), (\ref{contdet}),
and (\ref{Hcont}),
$\lim_{L\to\infty}\det (I+U_L)=\det_2(I+H)$. Thus from (\ref{limits})
and (\ref{detU}),  
 \begin{equation}
   R = \lim_{L\to\infty}[S(\hat\mu_L) - S(\hat\nu_L)]
  =   {1\over2}\log\det\nolimits_2(I + H).   \label{FD}
  \end{equation}

\noindent
{\bf Remark 4.1:} (a) It follows from (\ref{F1}) and (\ref{F2}) that
$|h(x,y|\le1$, with strict equality possible only for $x=y$ and
$\rho_a=1-\rho_b=1$; this implies
that the operator norm of $H$ is less than one, so that from (\ref{P3}) we
have the expansion
 \begin{equation}
R =\sum_{n=2}^\infty{(-1)^{n+1}\over 2n}\tr H^n.\label{Rsum}
 \end{equation}
 The convergence of the expansion (\ref{Rsum}) is the additional condition
for the validity of (\ref{limits2}) referred to in section~\ref{others}.
  
 \smallskip\noindent
 (b) The operators $-H_L$ may be shown to be positive semi-definite, and
hence $-H$ is also; from $\int_0^1(-h(x,x))\,dx<\infty$ and the continuity
of $h(x,y)$ it then follows that $H$ is of trace class \cite{Simon}.  Thus
in fact $\det_2(I+H)$ is given by (\ref{detdet2}).

 \section{Truncated correlations and entropy\label{probs}}

In this section we derive an expression for the probability of a
configuration, and an expansion for the entropy, in terms of truncated
correlation functions.  The results hold in a more general setting than the
specific models we are considering here.  Thus let $\Lambda$ be any finite
set and $\mu_\Lambda$ be a measure on the configurations $\tauul_\Lambda$ on
$\Lambda$: $\tauul_\Lambda=(\tau_i)_{i\in\Lambda}$, $\tau_i=0,1$.  In
particular, $\Lambda$ might be a subset of a larger set, say
$\Lambda\subset\bbz^d$, and $\mu$ the restriction of some measure on the
configurations on this larger set to the configurations on $\Lambda$.  For
any set $A=\{i_1,\ldots,i_k\}\subset\Lambda$ we will write $\mu_A$ for the
marginal of $\mu_\Lambda$ on configurations defined on $A$ and $\nu_A$ for
the product measure on the configurations on $A$ which has the same
one-site probabilities as does $\mu_\Lambda$: for a configuration
$\tauul_A=(\tau_{i_1},\ldots,\tau_{i_k})$ with $\tau_i=0,1$, and with
$t_i=\langle\tau_i\rangle_{\mu_\Lambda}$,
 \begin{equation}
\nu_A(\tauul_A)=\prod_{i\in A}t_i^{\tau_i}(1-t_i)^{1-\tau_i}
 =\prod_{i\in A}[\tau_it_i+(1-\tau_i)(1-t_i)].\label{nuLam}
\end{equation}

\subsection{Probabilities and truncated correlations\label{ptc}}

For the subset $A=\{i_1,\ldots,i_k\}\subset\Lambda$ we denote the truncated
correlation function
$\langle\tau_{i_1}\cdots\tau_{i_k}\rangle^T_{\mu_\Lambda}$ on the sites of
$A$ by $t_A$; if $A=\{i\}$ we usually write $t_i$ rather than $t_{\{i\}}$,
in accord with (\ref{nuLam}). Recall that $t_A$ is defined recursively by
writing the (untruncated) correlation function on the sites of $A$ as a
sum, over all partitions of $A$ into disjoint subsets, of the products of
the truncated functions for the subsets: letting $\P(A)$ denote the set of
all partitions of $A$ into disjoint subsets we have \cite{Ruelle}
 \begin{equation}
\langle\tau_{i_1}\cdots\tau_{i_k}\rangle_{\mu_\Lambda}
  = \sum_{\pi\in\P(A)}\prod_{B\in\pi}t_B,\label{trundef1}
 \end{equation}
 where $\pi$ labels a particular partition. 

 For use in the next subsection it is convenient to rewrite the measure
$\mu_\Lambda$ by factoring out the product measure $\nu_\Lambda$:
 \begin{equation}
\mu_\Lambda(\tauul_\Lambda)
   =\nu_\Lambda(\tauul_\Lambda)(1+x_\Lambda(\tauul_\Lambda)),\label{Px}
 \end{equation}
 where 
 \begin{equation}
  x_{\Lambda}(\tauul_\Lambda)= \sum_{\pi\in\Pt(\Lambda)}
  \prod_{i\in C_\pi}g_i(\tau_i)
    \prod_{B\in\pi}t_B. \label{xdef}
 \end{equation}
  Here $\Pt(A)$, $A\subset\Lambda$, denotes the set of nonempty
families $\pi=\{B_1,\ldots,B_{k(\pi)}\}$ of pairwise disjoint subsets of
$A$ in which each set $B_i$ contains at least two points, with
$C_\pi=\bigcup_iB_i$ for $\pi\in\Pt(A)$, and
 \begin{equation}
  g_i(\tau_i) = t_i^{-\tau_i}[-(1-t_i)]^{-(1-\tau_i)}
  = (-1)^{1-\tau_i}{1\over\nu_{\{i\}}(\tau_i)}.
 \label{pconfig}
 \end{equation}
 To verify this formula one multiplies both sides of (\ref{Px})  by some
product $\tau_{i_1}...\tau_{i_k}$ and sums over $\underline{\tau}_\Lambda$; 
the result is just  (5.2).

 \medskip\noindent
 {\bf Remark 5.1:} An alternate way of viewing (\ref{Px}) is to introduce
{\sl truncated measures} $\hat\mu_A$ defined by a recursion analogous to
that for the truncated correlation functions:
 \begin{equation}
 \mu_A(\tauul_A)
  = \sum_{\pi\in\P(A)}\prod_{B\in\pi}\hat\mu_B(\tauul_B).\label{tmdef}
 \end{equation}
 Disentangling (\ref{tmdef}), on sees that $\hat\mu_A$ is a linear
combination of measures on the configurations on $A$, but for $|A|>1$ with
some negative coefficients; that is, $\hat\mu_A$ is a signed measure.
There is a surprisingly simple relation between these truncated measures
and the truncated correlation functions: we claim that
 \begin{equation}
\hat\mu_{\{i\}}(\tau_i)=\mu_{\{i\}}(\tau_i),\label{tm1}
 \end{equation}
 and if $|B|\ge2$,
 \begin{equation}
\hat\mu_B(\tauul_B)=(-1)^{|B|}\prod_{i\in B}(-1)^{\tau_i} t_B
   =\biggl[\,\prod_{i\in B}(2\tau_i-1)\biggr] t_B;\label{tm2}
 \end{equation}
 i.e., $\hat\mu_B(\tauul_B)$ is equal to either $t_B$ or $-t_B$, depending
only on whether $\sum_{i\in B}(1-\tau_i)$ is even or odd.  Equation
(\ref{tm1}) is an immediate consequence of the definition (\ref{tmdef}).
Equation (\ref{tm2}) may be verified by substituting (\ref{tm1}) and
(\ref{tm2}) into the right hand side of (\ref{tmdef}), multiplying the
result by some product $\prod_{i\in C}\tau_i$, where $C\subset A$, and
summing over all $\tauul_A$; the result is just (\ref{trundef1}).  If we
now substitute (\ref{tm1}) and (\ref{tm2}) into  (\ref{tmdef}) we obtain
(\ref{Px}).  

\subsection{Expansion of the entropy\label{tce}}

In this subsection we obtain a series expansion for the entropy difference
$S(\mu_\Lambda) - S(\nu_\Lambda)$. Other graphical expansions for the
entropy have been obtained, for example in \cite{Stell,Bed}.  From
(\ref{Px}) and the definition (\ref{entropy}) we have
 \begin{eqnarray}
S(\mu_\Lambda)&=&-\sum_{\tauul_\Lambda}\nu_\Lambda(\tauul_\Lambda)
   (1+x_\Lambda(\tauul_\Lambda))
      \log[\nu_\Lambda(\tauul_\Lambda)(1+x_\Lambda(\tauul_\Lambda))]\nonumber\\
  &=& -\sum_{\tauul_\Lambda}\nu_\Lambda(\tauul_\Lambda)\Bigl
    [\log\nu_\Lambda(\tauul_\Lambda)
 +x_\Lambda(\tauul_\Lambda)\log\nu_\Lambda(\tauul_\Lambda)
     \nonumber\\
 &&\hskip30pt
  +(1+x_\Lambda(\tauul_\Lambda))\log(1+x_\Lambda(\tauul_\Lambda))\Bigr].
 \label{entropy17}
 \end{eqnarray}
 With the expansion
 \begin{equation}
(1+x)\log(1+x)= x 
   + \sum_{n=2}^\infty{(-1)^n\over n(n-1)}\,x^n,\label{sall}
 \end{equation}
 and the identities
 \begin{equation}
\sum_{\tauul_\Lambda}
\nu_\Lambda(\tauul_\Lambda)x_\Lambda(\tauul_\Lambda)=0,
 \qquad
\sum_{\tauul_\Lambda} \nu_\Lambda(\tauul_\Lambda)x_\Lambda(\tauul_\Lambda)
   \log\nu_\Lambda(\tauul_\Lambda)=0,
 \end{equation}
 which follow from (\ref{Px}) and the equations
$\langle1\rangle_{\mu_\Lambda}=\langle1\rangle_{\nu_\Lambda}=1$ and
$\langle\tau_i\rangle_{\mu_\Lambda}=\langle\tau_i\rangle_{\nu_\Lambda}=t_i$,
respectively, (\ref{entropy17}) yields
 \begin{equation}
 S(\mu_\Lambda)-S(\nu_\Lambda)
 =\sum_{\tauul_\Lambda}\nu_\Lambda(\tauul_\Lambda)
  \sum_{n=2}^\infty{(-1)^{n+1}\over n(n-1)}\,x_\Lambda(\tauul_\Lambda)^n.
 \label{entropy18}
 \end{equation}
 This expansion requires that $|x_\Lambda(\tauul_\Lambda)|<1$
for all $\tauul_\Lambda$; for the SSEP we have checked this condition
numerically for $L=1,\ldots,23$ at several values of $\rho_a$, $\rho_b$.

 We next insert the definition (\ref{xdef}) of $x_\Lambda$ into
(\ref{entropy18}) and expand $x_\Lambda(\tauul_\Lambda)^n$:
 \begin{equation}
 S(\mu_\Lambda)-S(\nu_\Lambda)
 =\sum_{\tauul_\Lambda}\nu_\Lambda(\tauul_\Lambda)
  \sum_{n=2}^\infty{(-1)^{n+1}\over n(n-1)}
  \sum_{\pi_1,\ldots\pi_n}\prod_{j=1}^n
   \left[\prod_{i\in C_{\pi_j}}g_i(\tau_i) \prod_{B\in\pi_j}t_B\right]
 \label{entropy19}
  \end{equation}
 This expression can be reorganized as an (infinite) linear combination of
monomials $M$ in the variables $t_B$. The coefficient of the monomial $M$
is
 \begin{equation}
 \sum_{n=2}^\infty{(-1)^{n+1}\over n(n-1)}\,c_n(M)
  \sum_{\tauul_\Lambda}\nu_\Lambda(\tauul_\Lambda)
     \prod_{i\in\Lambda}g_i(\tau_i)^{m_i(M)},
   \label{Mcoeff}
 \end{equation}
  where $c_n(M)$ is the number of $n$-tuples $(\pi_1,\ldots,\pi_n)$ such
that $M$ is given by the product $\prod_{j=1}^n\prod_{B\in\pi_j}t_B$, and
$m_i(M)$ is the number of factors $t_B$ in the monomial $M$ such that
$i\in B$.    

We can carry out the sum over $\tauul_\Lambda$ in (\ref{Mcoeff}), using (with
$m_i=m_i(M)$)
 \begin{equation}
   h_i(M) \equiv \sum_{\tau_i=0,1}t_i^{\tau_i}(1-t_i)^{1-\tau_i}
   g_i(\tau_i)^{m_i}
  =\left[{1\over t_i^{m_i-1}}+{(-1)^{m_i}\over(1-t_i)^{m_i-1}}
         \right].\label{hdef}
 \end{equation}
 Since $h_i(M)=1$ if $m_i(M)=0$, (\ref{entropy19}) and (\ref{Mcoeff})
yield
 \begin{equation}
 S(\mu_\Lambda)-S(\nu_\Lambda)
 = \sum_M  d(M)\,M \prod_{i\in D_M} h_i(M),\label{entropy2}
 \end{equation}
 where $D_M$ is the set of indices $i$ such that $m_i(M)>0$ and
 \begin{equation}
 d(M)= \sum_{n=2}^\infty{(-1)^{n+1}\over n(n-1)}\, c_n(M). \label{ddef}
 \end{equation}
 Since $h_i(M)=0$ if $m_i(M)=1$, we may restrict the sum in
(\ref{entropy2}) to monomials $M$ for which $m_i(M)\ge2$ for $i\in D_M$.

 We obtain a graphical representation for (\ref{entropy2}) by associating
to each monomial $M$ a graph $G_M$; $G_M$ has a vertex for each factor
$t_B$ in $M$, and vertices corresponding to factors $t_B,t_C$ are joined by
an edge if and only if $B\cap C\ne\emptyset$. We will show in Appendix B
that $d(M)=0$ unless $G_M$ is connected; from this observation together
with the remarks above it follows that the sum in (\ref{entropy2}) can be
restricted to the set $\M$ of all monomials for which $m_i(M)\ge2$ for all
$i\in D_M$ and for which $G_M$ is connected.  We also show in Appendix B
that if $G_M$ is a cycle with $k\ge3$ vertices then $d(M)=(-1)^{k+1}$.
Note finally that if $G_M$ consists of two vertices joined by an edge then
the requirement that $m_i(M)\ge2$ for $i\in D_M$ implies that $M=t_B^2$ for
some $B$, so that $c_n(M)=\delta_{n,2}$ and $d(M)=-1/2$ from (\ref{ddef}).

\section{Entropy for the SSEP\label{particles}}

We now apply the expansion (\ref{entropy2}) to the special case of the
SSEP, taking $\mu_\Lambda$ to be the NESS measure $\mubar_L$ on $\Lambda_L$
and thus $\nu_\Lambda$ to be the product measure $\nu_L^{(\bar\rho)}$,
which we will here write as $\bar\nu_L$.  We will use (\ref{Fk}) to
identify the order, as $L\to\infty$, of the terms in the series; we then
show that in this limit the leading order terms in this series sum to $R$.
We do not, however, give estimates which would completely justify the
neglect of the higher order terms.

Let us denote the order in $L$ of a monomial $M\in\M$ by $-j_M$, that is,
we suppose that $M$ is $O(L^{-j_M})$ as $L\to\infty$.  From (\ref{Fk}) we
know that $t_B$ is of order $L^{-(|B|-1)}$, so that if $M$ has $k$ factors
(not necessarily distinct), $M=t_{B_1}\cdots t_{B_k}$, then
$j_M=\sum_{i=1}^k (|B_i|-1)$.  Since $|B_l|\ge2$ for each $l$,
$k\le(1/2)\sum|B_l|$.  But then, because $m_i\ge2$ for $i\in D_M$,
 \begin{equation}
   j_M=\sum_{l=1}^k|B_l|-k\ge{1\over2}\sum_{l=1}^k|B_l|\ge|D_M|;\label{jm}
 \end{equation}
 note that $|D_M|$ is the total number of sites which belong to some $B_i$.
Equality holds in (\ref{jm}) if and only if
 \begin{equation}
\hbox{$|B_l|=2$ for each factor $t_{B_l}$ of $M$, and $m_i(M)=2$ for each
$i\in D_M$.}\label{lowdeg}
 \end{equation}
 The terms satisfying condition (\ref{lowdeg}) give the leading order
 contribution to $S(\mubar_L)-S(\nubar_L)$, as we now discuss.

   Let $\M_1\subset\M$ be the monomials for which $j_M=|D_M|$, that is,
those which satisfy (\ref{lowdeg}), and let $\M_2=\M\setminus\M_1$.  Then
we write (\ref{entropy2}) in the form 
 \begin{equation}
S(\mubar_L) - S(\nubar_L) = \sum_{k=2}^L
    \sum_{\ts{{A\subset\{1,\ldots,L\}\atop |A|=k}}}(s_{L,1}(A)+s_{L,2}(A)),
 \label{entropy3}
 \end{equation}
 where
 \begin{equation}
s_{L,j}(A) = \sum_{\ts{M\in\M_j\atop D_M=A}}
     d(M)\,M\prod_{i\in D_M} h_i(M), \qquad j=1,2,\label{slj}
 \end{equation}
 so that when $j=1$, each summand in (\ref{slj}) is of order
$L^{-|A|}$, while when $j=2$, each term is of higher order.  

  We first consider the sum of the $s_{L,1}(A)$.  For $M\in\M_1$,
(\ref{hdef}) and (\ref{lowdeg}) imply that $h_i(M)=1/[t_i(1-t_i)]$.  Moreover,
(\ref{lowdeg}) and the requirement that $G_M$ be connected imply that $G_M$
is a cycle or, if $|D_M|=2$, a single edge connecting two vertices, and for
these graphs we know the value of $d(M)$, as discussed at the end of
section~\ref{probs}.  This leads to
 \begin{equation}
  s_{L,1}(A) = {(-1)^{|A|+1}\over2}
   \sum_\sigma\prod_{i\in A} {t_{\{i,\sigma(i)\}}\over t_i(1-t_i)},
 \end{equation}
 where the sum is over all cyclic permutations $\sigma$ of $A$.  Here the
overall factor of 1/2 arises for $|A|=2$ from $d(M)=-1/2$ and for $|A|\ge3$
from the fact that the permutations from a cycle and from the reverse cycle
give rise to the same monomial.  Thus
 \begin{eqnarray}
  \sum_{|A|=k} s_{L,1}(A) 
  &=& {(-1)^{k+1}\over 2k}
   \sum_{1\le i_1\ne i_2\ne\cdots\ne i_k\le L} 
   (U_L)_{i_1i_2}(U_L)_{i_2i_3}\cdots (U_L)_{i_{k-1}i_k}(U_L)_{i_ki_1}
 \nonumber\\
 &=& {(-1)^{k+1}\over 2k}[\tr H_L^k + O(L^{-1})], \label{s1}
 \end{eqnarray}
 where $H_L$ was defined in (\ref{HLdef}) and the $O(L^{-1})$ error arises
from the fact that the sum omits terms in which some of the indices $i_j$
coincide.

 It follows from (\ref{s1}), (\ref{conttr}) and (\ref{Hcont}) that for
$k\ge2$,
 \begin{equation}
  \lim_{L\to\infty}\sum_{|A|=k} s_{L,1}(A) = {(-1)^{k+1}\over 2k}\tr H^k.
 \label{sl1lim}
 \end{equation}
 It is also true that
 \begin{equation}
  \lim_{L\to\infty}\sum_{k=2}^\infty\sum_{|A|=k} s_{L,1}(A) 
  = \sum_{k=2}^\infty{(-1)^{k+1}\over 2k}\tr H^k = R  \label{Rlim}
 \end{equation}
 (see (\ref{Rsum})); this will follow from (\ref{sl1lim}) and the Lebesgue
dominated convergence theorem if we show that
$\left|\sum_{|A|=k} s_{L,1}(A)\right| \le e_k$ for some convergent series
$\sum_ke_k$.  Now since $U_{ij}\le0$ for all $i,j$ and hence the
$O(L^{-1})$ term in (\ref{s1}) has the opposite sign to $\tr H_L^k$, and
since for any $\epsilon>0$,
$|h_L(x,y)|\le|h(x,y)|+\epsilon=|h(x,y)-\epsilon|$ for sufficiently large
$L$, we have from (\ref{trint}) that for such $L$,
 \begin{equation}
 \left|\sum_{|A|=k} s_{L,1}(A)\right| \le {1\over 2k}\Bigl|\tr H_L^k\Bigr|
   \le {1\over 2k}\Bigl|\tr (H-\epsilon C)^k\Bigr|,
 \label{s1bd}
 \end{equation}
 where $C$ is the integral operator with kernel $c(x,y)\equiv1$.  If we
take $\epsilon$ sufficiently small that $\|H-\epsilon C\|<1$ then
(\ref{s1bd}) furnishes the needed bound.

We now consider the contribution to (\ref{slj}) of the terms $s_{L,2}(A)$,
each of which is $O(L^{-j})$ with $j>|A|$.  Since there are order $L^k$
sets $A$ with $|A|=k$, we have formally that
 \begin{equation}
\lim_{L\to\infty}\sum_{|A|=k}s_{L,2}(A)=0.\label{formal}
 \end{equation}
 We will assume that (\ref{formal}) holds and  can in fact be extended to
 \begin{equation}
\lim_{L\to\infty}\sum_{k=2}^L\sum_{|A|=k}s_{L,2}(A)=0.\label{formal2}
 \end{equation}
 From (\ref{entropy3}), (\ref{Rlim}), and (\ref{formal2}) we have  
 \begin{equation}
   \lim_{L\to\infty}[S(\mubar_L)-S(\nubar_L)] 
 = R,
 \end{equation}
 which verifies (\ref{limits2}).

\section{Conclusion\label{conclude}}

In this paper we have seen how the truncated pair correlation function
which describes the Gaussian fluctuations of the density profile in the
non-equilibrium steady state of the simple exclusion process also
determines the leading correction, which is of order 1, to the entropy
$S(\mu)$.  One could also ask how the higher order truncated correlation
functions, which are related to higher order terms in the expansion of the
LDF around $\bar \rho$, contribute to further corrections to the entropy.
Going beyond the simple exclusion process (and the WASEP and KMP model), in
which local equilibrium corresponds to a product measure, it would be
interesting to consider more general non-equilibrium steady states in
which, in addition to the weak long range part of the correlation of the
form (1.1), there is an $O(1)$ short range part.  The simplest extension of
our result would be that the leading order term would again be given by
that of the local equilibrium and that the leading correction would again
be that coming from the non local part of the truncated pair correlation.
Another interesting extension would be to cases in which the leading order
of the entropy is still obtained from a local equilibrium product measure
but the long range part of the correlation obeys another scaling or in
which the fluctuations of the density are not Gaussian (as in the
asymmetric simple exclusion process; see \cite{DEL}).

For isolated systems at equilibrium, that is, in the microcanonical
ensemble, all microscopic configurations have equal probability, and so
$S(\mu)=-\log\mu=\log|\Omega|$, where $|\Omega|$ is the number of
configurations, or the phase space volume, available to the system.  When
one moves to the canonical ensemble, still at equilibrium, the
probabilities of configurations visited by the system fluctuate: the
Gibbs-Shannon entropy $S(\mu)$ is just the expectation of the logarithm of
these probabilities.  The variance of this logarithm is, up
to a trivial temperature factor, the variance of the energy; it is an
extensive quantity whose value per unit volume (or lattice site) $V(\mu)$ is
related to the specific heat.  One expects further that, in equilibrium
systems, the quantity $[-\log\mu-S(\mu)]/\sqrt{LV(\mu)}$ will in the
$L\to\infty$ limit approach a standard normal random variable; this is an
exercise for one-dimensional systems in \cite{normal}.  It is easily
verified that the same holds for the local equilibrium measure $\bar\nu_L$
considered here.

 A natural question now is: in what respect is the distribution of this
logarithm in nonequilibrium steady states, such as the NESS of the SSEP,
different from or similar to the distribution in equilibrium systems?  For
example, are there characteristics of this distribution which can be
related to physically measurable macroscopic quantities?

Although we do not know yet whether such questions have general answers, we
have measured for small system sizes the quantity $V(\mubar_L)$:
 \begin{equation}
V(\mubar_L) 
   = \frac{1}{L} \Bigl\langle[- \log \mubar_L(\tau_L) - S(\mubar_L)]^2
    \Bigr\rangle_{\mubar_L}. \label{VmuL}
 \end{equation}
  Our results are plotted in Figure 4 for $\rho_a=1$ and $\rho_b=0$;
$V(\mubar_L)$ appears to approach a fixed value in the large L limit. For
comparison we have also plotted there the corresponding quantity
$V^*(\mubar_L)$ defined by
 \begin{equation}
V^*(\mubar_L) 
   = \frac{1}{L} \Bigl\langle[- \log \bar\nu_L(\tau_L) - S(\nubar_L)]^2
    \Bigr\rangle_{\mubar_L}. \label{VmuLstar}
 \end{equation}
 \begin{figure} [!t]
\centerline{\epsffile{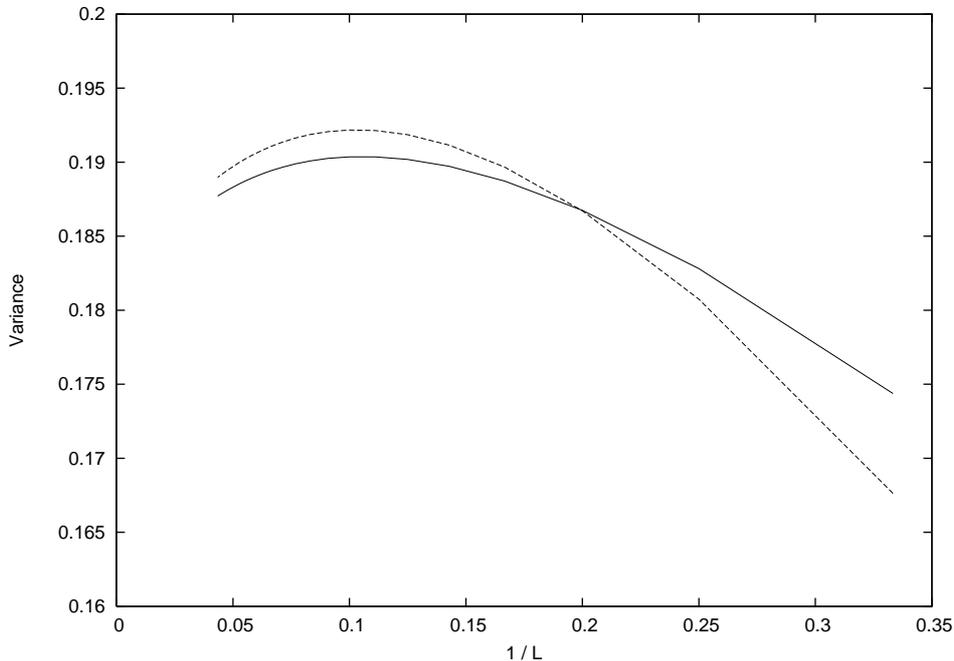}}
 \caption{Variances $V(\mubar_L)$ (solid line) 
and $V^*(\mubar_L)$ (dashed line), evaluated for 
small systems for the SSEP with $\rho_a=1$ and $\rho_b=0$,
plotted against $1/L$. Equation (\ref{Vinf})
 predicts a large $L$ convergence to 0.179956...} 
\end{figure}%
 In fact, by repeating the arguments of sections 5 and 6 one can show that
the limiting values of these two quantities coincide and are given by
 \begin{eqnarray}
  \lim_{L\to\infty}V(\mubar_L) &=& 
     \int_0^1dx\,F_1(x)(1-F_1(x))\left(\log\frac{F_1(x)}{1-F_1(x)}\right)^2+
  \nonumber\\
  &&\hskip-60pt
 2\int_0^1dx\,\int_x^1dy\,
  F_2(x,y)\left(\log\frac{F_1(x)}{1-F_1(x)}\right)
   \left(\log\frac{F_1(y)}{1-F_1(y)}\right)\label{Vinf}.
 \end{eqnarray}
 The first term in (\ref{Vinf}) is the corresponding quantity for the local
 equilibrium system: 
 \begin{eqnarray}
  \lim_{L\to\infty}V(\bar\nu_L) 
  &=&  \lim_{L\to\infty}
  \frac{1}{L} \Bigl\langle[- \log \bar\nu_L(\tau_L) - S(\nubar_L)]^2
    \Bigr\rangle_{\bar\nu_L}\nonumber\\ 
    &=&
    \int_0^1dx\,F_1(x)(1-F_1(x))\left(\log\frac{F_1(x)}{1-F_1(x)}\right)^2.
    \label{Vleinf}
 \end{eqnarray}
  The difference between (\ref{Vinf}) and (\ref{Vleinf}) shows that in
contrast to the entropy itself, for which local equilibrium gives correctly
the leading order in $L$ \cite{Bahadoran}, the two point correlations
affect the leading order of the variance $LV(\bar \mu_L)$.
For $\rho_a=1$ and $\rho_b=0$ the expression (\ref{Vinf}) takes the value
$\pi^2/9-11/12\approx 0.179956$, and the expression (\ref{Vleinf}) the
value $(\pi^2-6)/18 \approx 0.214978$.

\bigskip\noindent
{\bf Acknowledgments:} We thank C.~Bahadoran, S.~Goldstein, and O.~Penrose
for helpful discussions, and C.~Bahadoran for sending us a copy of
\cite{Bahadoran}.  J.L.L. and E.R.S acknowledge the hospitality of the
I.H.E.S.~in the spring of 2005.  The work of J.L.L was supported in part by
NSF Grant DMR--0442066 and AFOSR Grant F49620; any opinions, findings and
conclusions, or recommendations expressed in this material are those of the
authors, and do not necessarily reflect the views of the NSF.  B.D. thanks 
the ACI-NIM 168 Transport Hors Equilibre of the Minist\`ere de l'Education
Nationale for support.

\appendix

\section{Correlation functions in the SSEP\label{ssep}}

Correlation functions in the SSEP may be obtained via the {\sl matrix
method} \cite{DEHP}. One introduces matrices $D$ and $E$ and vectors
$|V\rangle$ and $\langle W|$ which satisfy
 \begin{eqnarray}
     DE - ED &=& D + E  \label{DE}\;,\label{alg1}\\
  (\beta D - \delta E)|V\rangle &=& |V\rangle\;,\label{alg2}\\
   \langle W|(\alpha E - \gamma D) &=& \langle W|\;,\label{alg3} 
 \end{eqnarray}
 where $\alpha$, $\beta$, $\gamma$, and $\delta$ were defined in
 section~\ref{models}.  Then 
 \begin{equation}
 \mubar_L(\tau_1,\ldots,\tau_L)=  
  {\langle W|(\tau_1D+(1-\tau_1)E)\cdots(\tau_LD+(1-\tau_L)E)|V\rangle
    \over  \langle W|(D+E)^L|V\rangle}\;, \label{prob}
 \end{equation}
 and so (in this section we write
$\langle\cdot\rangle_{\mubar_L}\equiv\langle\cdot\rangle_L$)
 \begin{equation}
   \langle\tau_{i_1}\cdots\tau_{i_k}\rangle_L = 
  {\langle W|(D+E)^{i_1-1}D(D+E)^{i_2-i_1-1}D\cdots D(D+E)^{L-i_k}|V\rangle
    \over  \langle W|(D+E)^L|V\rangle}\;. \label{exp}
 \end{equation}
 The normalization factor in (\ref{exp}) has been evaluated in \cite{LDF}:
 \begin{equation}
 \langle W|(D+E)^L|V\rangle
  ={\Gamma(a+b+L)\over\Gamma(a+b)(\rho_a-\rho_b)^L}\langle W|V\rangle.
 \end{equation}
  Now one obtains a recursion relation for the correlation functions:
starting from the formula (\ref{exp}) for
$\langle\tau_{i_1}\cdots\tau_{i_k}\tau_{i_{k+1}}\rangle_L$, one first
commutes the rightmost factor of $D$ to the extreme right in the product,
using $[D,D+E]=D+E$, then writes
$D|V\rangle=(\beta+\delta)^{-1}\bigl(|V\rangle+\delta(D+E)|V\rangle\bigr)$
(which is equivalent to (\ref{alg2})); the result is
 \begin{eqnarray}
 \langle\tau_{i_1}\cdots\tau_{i_k}\tau_{i_{k+1}}\rangle_L 
 &=&  {(\rho_a-\rho_b)(L+b-i_{k+1})\over L+a+b-1}
   \langle\tau_{i_1}\cdots\tau_{i_k}\rangle_{L-1} \nonumber\\
  &&\hskip40pt
  +\rho_b\langle\tau_{i_1}\cdots\tau_{i_k}\rangle_L. \label{recur0}
  \end{eqnarray}
  Taking $k=0$, $1$, and $2$ one recovers (\ref{tau1}), (\ref{tau2}), and
(\ref{tau3});  (\ref{recur0}) may now be written in the form
 \begin{equation}
 \langle\tau_{i_1}\cdots\tau_{i_k}\tau_{i_{k+1}}\rangle_L
  = (\langle\tau_{i_{k+1}}\rangle_L-\rho_b)
     \bigl(\Delta\langle\tau_{i_1}\cdots\tau_{i_k}\rangle\bigr)_L
  +\langle\tau_{i_{k+1}}\rangle_L\langle\tau_{i_1}\cdots
            \tau_{i_k}\rangle_L,
  \;\;\;\label{bdrecur}
 \end{equation}
  where for any sequence $c_1,c_2,c_3,\ldots$ we write
 \begin{equation}
  (\Delta c)_L=c_{L-1}-c_L, \qquad L\ge2.\label{Delta}
 \end{equation}

The truncated correlation functions
$t_{A,L}\equiv\langle\tau_{i_1}\cdots\tau_{i_k}\rangle^T_L$, where
$A=\{i_1,\ldots,i_k\}$ with $k\ge1$, are defined recursively by
 \begin{equation}
\langle\tau_{i_1}\cdots\tau_{i_k}\rangle_L
  = \sum_{\pi\in\P(A)}\prod_{B\in\pi}t_{B,L},\label{trundef}
 \end{equation}
 (see (\ref{trundef1})). We claim that for $k\ge1$ these functions satisfy
the recursion
 \begin{equation}
 t_{A\cup\{i_{k+1}\},L}
     =(\langle\tau_{i_{k+1}}\rangle_L-\rho_b)
    \sum_{\pi\in\P(A)}\prod_{B\in\pi} (\Delta t_B)_L,\label{recur}
 \end{equation}
 which, together with $t_{\{i\},L}=\langle\tau_i\rangle_L$, determines all
the $t_{A,L}$.  We will verify (\ref{recur}) below, after we have shown
that it implies (\ref{Fk}).

It follows from (\ref{recur}) that for $A=\{i_1,\ldots,i_k\}$,
$t_{A,L}=v_L^k(\iul)$, where $\iul=(i_1,\ldots,i_k)$ and $v^k_L(\iul)$ is a
rational function of $L$ and $i_1,\ldots,i_k$ which is a polynomial of
degree 1 in each of the $i_j$.  For $\xul=(x_1,\ldots,x_k)\in\bbr^k$ let us
define $u_L^k(\xul)=v_L^k(L\xul)$; $u$ is again rational and a polynomial
of degree 1 in each $x_j$, so that we will obtain (\ref{Fk}) if we show
that $u_L^k=O(L^{-k+1})$.  We show this by induction on $k$; for $k=1$ it
is an immediate consequence of (\ref{tau1}).  But if 
$u_L^k=O(L^{-k+1})$ for $k<k_0$ then from 
 \begin{equation}
(\Delta v^k)_L(L\xul) = (\Delta u^k)_L(\xul) 
   + \bigl[u^k_{L-1}([1+(L-1)^{-1}]\xul)-u^k_{L-1}(\xul)\bigr]
 \end{equation}
 it follows that $(\Delta v^k)_L(L\xul)=O(L^{-k})$ if $k<k_0$, and
$u_L^k=O(L^{-k+1})$ for $k=k_0$ follows by evaluating (\ref{recur}) at
$\iul=L\xul$.

\medskip\noindent
 {\bf Remark A.1:} The recursion (\ref{recur}) implies a similar recursion for
the $F_k$.  Recall that the operator $\sum_{i=1}^kx_i\partial/\partial x_i$
acts on a monomial of degree $d$ in $x_1,\ldots,x_k$ as multiplication by
$d$, so that the operator $D_k=k-1+\sum_{i=1}^kx_i\partial/\partial x_i$
multiplies such a monomial by $k+d-1$.  Then
 \begin{equation}
 F_{k+1}(x_1,\ldots,x_k,x_{k+1})
     =(F_1(x_{k+1})-\rho_b)
    \sum_{\pi\in\P(\{1,\ldots,k\})}
    \prod_{B\in\pi}[ D_{|B|}F_{|B|}] ((x_i)_{i\in B}).\label{recurF}
 \end{equation}
 This follows by writing $t_{A,L}=\sum_d L^{-(k+d-1)}P_d+\rm h.o.t.$, where
$P_d$ is homogeneous of degree $d$ in $i_1,\ldots,i_k$ and h.o.t.~denotes
terms which are $O(L^{-k})$ after the substitutions $i_j=Lx_j$,
$j=1,\ldots,L$.  We will not use this formula and so omit further details,
but we do note that an easy consequence is that, for $k\ge2$, $F_k$ depends
on $\alpha$, $\beta$, $\gamma$, and $\delta$ only through an overall factor
of $(\rho_a-\rho_b)^k$.

 \medskip\noindent
{\bf Proof of the recursion (\ref{recur}):} To verify (\ref{recur}) we need
a formula for the action of $\Delta$ on a product (see (\ref{Delta})).
Suppose that $c^{(1)},\ldots,c^{(k)}$ are sequences (i.e.,
$c^{(i)}=(c^{(i)}_L)_{L=1}^\infty$) and that we multiply such sequences
componentwise, so that
$(c^{(1)}\cdots c^{(k)})_L= c^{(1)}_L\cdots c^{(k)}_L$.  Then trivially
 \begin{equation}
(1+\Delta)(c^{(1)}\cdots c^{(k)})=[(1+\Delta)c^{(1)}]\cdots
   [(1+\Delta)c^{(k)}],
 \end{equation}
  and so with $X=\{1,2,\ldots,k\}$,
 \begin{equation}
 \Delta(c^{(1)}\cdots c^{(k)}) 
  = \sum_{\emptyset\ne Y\subset X}\prod_{i\in Y}\Delta c^{(i)}
   \prod_{j\in X\setminus Y}c^{(j)}.\label{id}
 \end{equation}
 For example,
 \begin{eqnarray}
 \Delta(c^{(1)}c^{(2)})
    &=&\Delta c^{(1)}c^{(2)}+c^{(1)}\Delta c^{(2)}
    +\Delta c^{(1)}\Delta c^{(2)}, \label{id2}\\
 \Delta(c^{(1)}c^{(2)}c^{(3)})
    &=&\Delta c^{(1)}c^{(2)}c^{(3)}+ c^{(1)}\Delta c^{(2)}c^{(3)}
    +c^{(1)}c^{(2)}\Delta c^{(3)}\nonumber\\
    &&\hskip20pt+\Delta c^{(1)}\Delta c^{(2)}c^{(3)}
     +\Delta c^{(1)}c^{(2)}\Delta c^{(3)}
    +c^{(1)}\Delta c^{(2)}\Delta c^{(3)}\nonumber\\
    &&\hskip20pt +\Delta c^{(1)}\Delta c^{(2)}\Delta c^{(3)}. \label{id3}
 \end{eqnarray}

We now verify (\ref{recur}); the case $k=1$ is precisely (\ref{bdrecur})
for $k=1$, and we proceed by induction on $k$.  We use (\ref{trundef})
to write $\langle\tau_{i_1}\cdots\tau_{i_k}\tau_{i_{k+1}}\rangle_L$, the
left hand side of (\ref{bdrecur}), in terms of truncated correlations,
separating the terms in which $i_{k+1}$ is grouped with some element of a
partition of $A$ from those in which $\{i_{k+1}\}$ is an element of the
partition of $A\cup\{i_{k+1}\}$:
 \begin{equation}
 \sum_{\pi\in\P(A)}\sum_{B\in\pi}t_{B\cup\{i_{k+1}\},L}
    \prod_{\ts{C\in\pi\atop C\ne B}}t_{C,L}
 +t_{\{i_{k+1}\},L}\sum_{\pi\in\P(A)}\prod_{B\in\pi}t_{B,L}.\label{bdl}
 \end{equation}
 On the other hand, with (\ref{trundef}) the right hand side of
(\ref{bdrecur}) becomes
 \begin{equation}
 (\langle\tau_{i_{k+1}}\rangle_L-\rho_b)
     \sum_{\pi\in\P(A)} (\Delta\prod_{B\in\pi}t_B)_L
  +\langle\tau_{i_{k+1}}\rangle_L \sum_{\pi\in\P(A)} \prod_{B\in\pi}t_{B,L}.
 \label{bdr}
 \end{equation}
 Since (\ref{bdl}) and (\ref{bdr}) are the two sides of (\ref{bdrecur}) we
 may equate these expressions to obtain
 \begin{equation}
   \sum_{\pi\in\P(A)}\sum_{B\in\pi}t_{B\cup\{i_{k+1}\},L}
    \prod_{\ts{C\in\pi\atop C\ne B}}t_{C,L}
  = (\langle\tau_{i_{k+1}}\rangle_L-\rho_b)
     \sum_{\pi\in\P(A)} (\Delta\prod_{B\in\pi}t_B)_L. \label{step}
 \end{equation}
 On the right hand side of (\ref{step}) we use (\ref{id}) to write
 \begin{equation}
  \sum_{\pi\in\P(A)}(\Delta\prod_{B\in\pi}t_B)_L
  =\sum_{\pi\in\P(A)}\sum_{\emptyset\ne\sigma\subset\pi}
   \prod_{C\in\sigma}(\Delta t_C)_L 
   \prod_{C\in\pi\setminus\sigma}t_{C,L}.
 \end{equation}
 On the left side of (\ref{step}) the term with $\pi=\{A\}$ is just
$t_{A\cup\{i_{k+1}\},L}$; we take the remaining terms to the other side of
the equation and in these terms use the induction assumption to write
 \begin{equation}
  t_{B\cup\{i_{k+1}\},L} = (\langle\tau_{i_{k+1}}\rangle_L-\rho_b)
  \sum_{\sigma\in\P(B)}\prod_{C\in\sigma} (\Delta t_C)_L.
 \end{equation}
  After these manipulations, (\ref{step}) becomes
 \begin{eqnarray}
 t_{A\cup\{i_{k+1}\},L}&=&(\langle\tau_{i_{k+1}}\rangle_L-\rho_b)
  \Big[\sum_{\pi\in\P(A)}\sum_{\emptyset\ne\sigma\subset\pi}
   \prod_{C\in\sigma}(\Delta t_C)_L 
   \prod_{C\in\pi\setminus\sigma}t_{C,L}\nonumber\\
  &&\hskip20pt
   -\sum_{\ts{\pi\in\P(A)\atop \pi\ne\{A\}}}
\sum_{B\in\pi}\sum_{\sigma\in\P(B)}\prod_{C\in\sigma} 
   (\Delta t_C)_L\prod_{\ts{C\in\pi\atop C\ne B}}t_{C,L}\Bigr].\ \ \ 
 \end{eqnarray}

 We now reorganize this expression. In the first sum we separate the term
$\pi=\{A\}$, which is simply $(\Delta t_A)_L$ (since necessarily
$\sigma=\{A\}$ also); in the remaining terms of this sum we relabel $\pi$
as $\pi'$, with $\pi'\ne\{A\}$.  Now every term in the second sum is
labeled by a partition $\pi$, a distinguished set $B\in\pi$, and a further
partition $\sigma$ of $B$; this data clearly gives rise to a new partition
$\pi'$ of $A$, $\pi'=(\pi\cup\sigma)\setminus\{B\}$, and a distinguished
subset $\sigma$ of $\pi'$; note that $\sigma\ne\emptyset$ since $\sigma$ is
a partition of $B$ and $\sigma\ne\pi'$ since $\pi\ne\{A\}$ and hence
$|\pi|\ge2$.  Thus
 \begin{eqnarray}
 t_{A\cup\{i_{k+1}\},L}
 &=& (\langle\tau_{i_{k+1}}\rangle_L-\rho_b)
  \Bigl[(\Delta t_A)_L
      \nonumber\\
  &&\hskip20pt
  +\sum_{\ts{\pi'\in\P(A)\atop \pi'\ne\{A\}}}\Bigl(
   \sum_{\emptyset\ne\sigma\subset\pi'}
   \prod_{C\in\sigma}(\Delta t_C)_L\prod_{C\in\pi'\setminus\sigma}t_{C,L}
   \nonumber\\
  &&\hskip60pt
  -\sum_{\emptyset\ne\sigma\subsetneq\pi'}
    \prod_{C\in\sigma}(\Delta t_C)_L\prod_{C\in\pi'\setminus\sigma}t_{C,L}
     \Bigr)\Bigr].
 \end{eqnarray}
 In the sum over $\pi'$ only the terms with $\sigma=\pi'$ survive, leading
to (\ref{recur}):
 \begin{equation}
  t_{A\cup\{i_{k+1}\},L}= (\langle\tau_{i_{k+1}}\rangle_L-\rho_b)
 \sum_{\pi'\in\P(A)}\prod_{B\in\pi'} (\Delta t_B)_L.
 \end{equation}

\section{The coefficients $d(M)$}

In this appendix we derive the two properties of the combinatorial factors
$d(M)$ (see (\ref{ddef})) which are needed in section~\ref{particles}: that
$d(M)=0$ if $G_M$ is not connected, and that $d(M)=(-1)^{k+1}$ if $G_M$ is
a cycle on $k$ vertices.  Our approach is to relate $d(M)$ to the number of
colorings of the graph $G_M$.  The condition that only monomials $M$ for
which $m_i(M)\ge2$ for all $i\in D_M$ occur in (\ref{entropy2}) implies
that every component of $G_M$ contains at least two vertices, and we assume
that all graphs considered in what follows satisfy this condition.

For any graph $G$ we let $\bar c_n(G)$ be the number of $n$-colorings
of $G$, where an {\sl $n$-coloring} of $G$ is an assignment of colors to
the vertices of $G$, using exactly $n$ colors, in such a way that adjacent
vertices are given distinct colors.  For example, if $G$ is a cycle on four
vertices then $\bar c_4(G)=24$, $\bar c_3(G)=12$, and $\bar c_2(G)=2$. Note
that the condition that every component of $G$ have at least two vertices
implies that $\bar c_1(G)=0$.  We also define
 \begin{equation}
 \bar d(G)= \sum_{n=2}^\infty{(-1)^{n+1}\over n(n-1)}\bar c_n(G).
      \label{dbardef}
 \end{equation}

Suppose now that $M$ is a monomial occurring in (\ref{entropy2}), say
$M=\prod_Bt_B^{k_B}$ with the $t_B$ distinct; the graph $G_M$ has
$\sum_Bk_B$ vertices, and we will denote the $k_B$ vertices corresponding
to $B$ by $v_{B,1},\ldots,v_{B,k_B}$.  Recall that $c_n(M)$ is the number
of $n$-tuples $\piul=(\pi_1,\ldots,\pi_n)$, with $\pi_i\in \Pt(D_M)$, such
that $M=\prod_{j=1}^n\prod_{C\in\pi_j}t_C$.  An $n$-coloring of $G_M$
immediately yields such a $\piul$, by taking $\pi_i$ to be the set of all
$B$ such that some $v_{B,j}$ is assigned color $i$.  Each $\piul$ arises in
this way from $\prod_Bk_B!$ distinct colorings, since for each $B$ we may
permute the colors assigned to the $v_{B,j}$ without changing $\piul$.
Thus $\bar c_n(G_M)=c_n(M)\prod_Bk_B!$, and 
 \begin{equation}
 \bar d(G_M)=d(M)\prod_B k_B!\;. \label{ddbar}
 \end{equation}

We next derive a recursion relation for $\bar d(G)$.  We first select some
vertex $v$ of $G$, and let $N_v$ be the set of vertices of $G$ which are
adjacent to $v$.  Every $n$-coloring of $G$ induces a partition $\lambda$
of $N_v$, where two vertices are in the same set of the partition iff they
have the same color; note that vertices in $N_v$ which are adjacent cannot
lie in the same element of $\lambda$.  Conversely, given any partition
$\lambda$ of $N_v$ satisfying this latter restriction we define the graph
$G_\lambda$ by (i)~removing from $G$ the vertex $v$ and all edges adjacent
to it, (ii)~collapsing all vertices belonging to a single subset
$B\in\lambda$ into a single vertex $w_B$ in $G_\lambda$ (which may give a
multi-graph; if so, we replace any multiple edges by a single edge), and
(iii)~joining each pair $w_B,w_{B'}$ of new vertices produced in this way
by an edge.  Then every $n$-coloring of $G$ is obtained by choosing
$\lambda$ and then either (a)~choosing one of the $n$ colors to assign to
$v$ and using the remaining colors for some $(n-1)$-coloring of
$G_\lambda$, or (b)~choosing an $n$-coloring of $G_\lambda$, then choosing
one of the $n-|\lambda|$ colors not used on the new vertices $w_B$ to
assign to $v$. This leads to the recursion
 \begin{equation}
 \bar c_n(G)
  =\sum_\lambda[n\bar c_{n-1}(G_\lambda)+(n-|\lambda|)\bar c_n(G_\lambda)].
 \end{equation}
 Then if no $G_\lambda$ is 1-colorable, so that $\bar c_{n-1}(G_\lambda)=0$
 if $n=2$, 
 \begin{eqnarray}
 \bar d(G)
 &=& \sum_{n=2}^\infty{(-1)^{n+1}\over n(n-1)}\,\bar c_n(G)\nonumber\\
 &=&\sum_\lambda\Bigl[
      \sum_{n=2}^\infty{(-1)^{n+1}\over (n-1)(n-2)}(n-2)\bar c_{n-1}(G_\lambda)
  \nonumber\\
 &&\hskip40pt +\sum_{n=2}^\infty{(-1)^{n+1}\over n(n-1)}
       (n-|\lambda|)\bar c_n(G_\lambda)\Bigr]  \nonumber\\
 &=&\sum_\lambda \sum_{n=2}^\infty{(-1)^{n+1}\over n(n-1)}
   [n-|\lambda|-(n-1)]\bar c_n(G_\lambda)\nonumber\\
 &=&\sum_\lambda (1-|\lambda|)\bar d(G_\lambda).\label{Grecur}
 \end{eqnarray}
 This is the desired recursion.

As a first consequence of (\ref{Grecur}) we show that if $G$ is a
disconnected graph in which every component has at least two
vertices, then $\bar d(G)=0$.  We argue by induction on the number $n$ of
vertices of $G$; certainly $n\ge4$.  If $n=4$ then $G$ has two components,
each a single edge joining two vertices, and an application of
(\ref{Grecur}) shows that $\bar d(G)=0$ (there will be only one partition
$\lambda$, with $|\lambda|=1$, in the sum).  We now argue by
induction on $n$; if we apply (\ref{Grecur}) with any vertex $v$ of $G$, every
$\bar d(G_\lambda)$ on the right hand side will vanish by the induction
assumption unless the ``new'' component of $G_\lambda$ has a single vertex,
in which case $|\lambda|=1$; thus $\bar d(G)=0$.

 As a second application we compute $\bar d(G)$ for $G$ a cycle.  First
note that if $G$ is a graph with 2 vertices joined by an edge then
$\bar d(G)=-1$, by a simple direct calculation.  If $G$ is a cycle with
$k\ge 3$ vertices and $v$ is any vertex of $G$ then $N_v$ contains two
vertices, say $N_v=\{w_1,w_2\}$, and the sum in
(\ref{Grecur}) has one term $\lambda=\lambda_0\equiv\{\{w_1\},\{w_2\}\}$
and, if $k\ge4$ so that $w_1$ and $w_2$ are not adjacent, also one with
$\lambda=\{N_v\}$.  The latter term, even if present, does not contribute
since $|\lambda|=1$, so $\bar d(G)=-\bar d(G_{\lambda_0})$.  But
$G_{\lambda_0}$ is a cycle with $k-1$ vertices or, if $k=3$, the two-vertex
graph considered above; thus $\bar d(G)=(-1)^{k+1}$ by induction on $k$.


\begin{thebibliography}{99}

\bibitem{KMP} C.~Kipnis, C.~Marchioro, and E.~Presutti,
 Heat-flow in an exactly solvable model, 
 {\sl J.~Stat.~Phys} {\bf 27}, 65--74 (1982). 

\bibitem{DFIP} A. De Masi, P. Ferrari, N. Ianiro, and E. Presutti, Small
deviations from local equilibrium for a process which exhibits
hydrodynamical behavior I, II, {\it J.~Stat.~Phys.}~{\bf 29} 57--79,
81--93 (1982).

 \bibitem{S2} H. Spohn, Long-range correlations for stochastic lattice gases
in a non-equilibrium steady-state, {\sl Journal of Physics}~{\bf A16},
4275-4291 (1983).

\bibitem{DEHP} B.~Derrida, M.~R.~Evans, V.~Hakim, V.~Pasquier,
Exact solution of a 1D asymmetric exclusion model using a matrix
formulation, {\sl J. Phys.~A}~{\bf 26},  1493--1517 (1993).

\bibitem{SD} G.~Schutz and R.~Domany, Phase-transitions in an exactly
soluble one-dimensional exclusion process, J.~Stat.~Phys {\bf 72}, 277-296
(1993).
 
\bibitem{DLS0} B.~Derrida, J.~L.~Lebowitz, and E.~R.~Speer, Free energy
functional for nonequilibrium systems: An exactly solvable case,
Phys.~Rev.~Lett~{\bf87}, 150601 (2001).


 \bibitem{BDGJL} L.~Bertini, A.~De Sole, D.~Gabrielli, G.~Jona-Lasinio,
C.~Landim, Fluctuations in stationary non equilibrium states of
irreversible processes, {\sl Phys.~ Rev.~ Lett.}~{\bf 87}, 040601 (2001). 

\bibitem{D} A. Dhar, Heat conduction in a one-dimensional gas of
elastically colliding particles of unequal masses, {\sl Phys.~Rev.~Lett}
{\bf 86}, 3554--3557 (2001).

 \bibitem{BDGJL2} L.~Bertini, A.~De Sole, D.~Gabrielli, G.~Jona-Lasinio,
C.~Landim, Macroscopic fluctuation theory for stationary non equilibrium
states, {\sl J. Stat.~Phys.}~{\bf 107}, 635--675 (2002).

\bibitem{LDF} B.~Derrida, J.~L.~Lebowitz, and E.~R.~Speer, Large deviation
of the density profile in the steady state of the open symmetric simple
exclusion process, {\sl J.~Stat.~Phys.}~{\bf 107}, 599--634 (2002).

\bibitem{JP} V. Jaksic, and C.-A. Pillet, Non-equilibrium steady states of
finite quantum systems coupled to thermal reservoirs, {\sl
Commun.~Math.~Phys.} {\bf 226}, 131--162 (2002).

\bibitem{DLS2} B.~Derrida, J.~L.~Lebowitz, and E.~R.~Speer, Exact large
deviation functional of a stationary open driven diffusive system: the
asymmetric exclusion process, {\sl J.~Stat.~Phys.}~{\bf 110}, 775--810 (2003).

\bibitem{LLP} S.~Lepri, R. Livi, and A. Politi, Thermal conduction in
classical low-dimensional lattices , {\sl Phys.~Reports} {\bf 377}, 1--80
(2003).

\bibitem{ED} C. Enaud and B. Derrida, Large deviation functional of the weakly
asymmetric exclusion process, {\sl  J.~Stat.~Phys.}~{\bf 114},
537--562 (2004).
    
\bibitem{EY}  J.-P. Eckmann and L.-S.~Young,
Temperature profiles in Hamiltonian heat conduction ,
{\sl Europhys.~Lett.} {\bf 68},  790--796  (2004).

 \bibitem{S} H.~Spohn, {\sl Large Scale Dynamics of Interacting Particles}
(Springer-Verlag, Berlin, 1991).

\bibitem{Bahadoran} C. Bahadoran, On the convergence of entropy for
  stationary exclusion processes with open boundaries, preprint 2004. 

\bibitem{Var} S.~R.~S.~Varadhan, Large Deviations and Entropy, in {\sl
  Entropy}, ed.~A. Greven, G. Keller, and G. Warnecke, Princeton University
  Press, Princeton, 2003.

 \bibitem{Olla} S. Olla, Large deviations for Gibbs random fields, {\sl
Probab.~Th.~Rel. Fields}~{\bf 77}, 343--357 (1988).

 \bibitem{Ellis} R. Ellis, {\sl Entropy, large deviations, and statistical
mechanics} (Springer, New York, 1985).

\bibitem{Kosygina} E. Kosygina, The behavior of the specific entropy in the
  hydrodynamic scaling limit, {\sl Ann.~Prob.}~{\bf 29}, 1086--1110 (2001).

\bibitem{WASEP} B. Derrida, C. Enaud, C. Landim, and S. Olla, Fluctuations
in the weakly asymmetric exclusion process with open boundary conditions,
{\sl J.~Stat.~Phys} {\bf 118}, 795--811 (2005).

\bibitem{BGL} L.~Bertini, D.~Gabrielli, and J.~L.~Lebowitz, Large
deviations for a stochastic model of heat flow. {\it
  J.~Stat.~Phys.}~{\bf121}: 843-885 (2005).

\bibitem{Simon} B. Simon, {\sl Trace Ideals and Their Applications},
Cambridge University Press, Cambridge, 1979.

\bibitem{Ruelle} D. Ruelle, {\sl Statistical Mechanics: Rigorous Results},
  W.~A.~Benjamin, New York, 1974.

\bibitem{Stell} G.~Stell, Cluster expansions for classical systems in
equilibrium, in {\sl The Equilibrium Theory of Classical Fluids}, ed.
H.~Frisch and J.~L.~L. Lebowitz, W.~A.~Benjamin, New York, 1964, and
references therein.

\bibitem{Bed} A. Bednorz, Graphical representation of the excess entropy,
{\it Physica A} {\bf 298}, 400-418 (2001).

\bibitem{DEL} B.~Derrida, C.~Enaud, and J.~L.~Lebowitz, 
The asymmetric exclusion process and Brownian
excursions, {\sl J.~Stat.~Phys.} {\bf 115}, 365--382 (2004).

\bibitem{normal} D. Ruelle, {\sl Thermodynamic Formalism,} Addison-Wesley,
  Reading, 1978.



\end{thebibliography}
 \end{document}